\documentclass[draft]{journal2019}
\usepackage{url} 
\usepackage{lineno}
\usepackage[inline]{trackchanges} 
\usepackage{soul}
\draftfalse

\journalname{JGR - Space Physics}

\begin{document}

\title{Exploring the origin of multi-periodic pulsations during a white-light flare}

\authors{Dong~Li\affil{1,2}\thanks{lidong@pmo.ac.cn}, Ding~Yuan\affil{3}, Jingye~Yan\affil{2}, Xinhua~Zhao\affil{2}, Zhao~Wu\affil{4}, Jincheng~Wang\affil{5,6}, Zhenyong~Hou\affil{7}, Chuan~Li\affil{8}, Haisheng~Zhao\affil{9}, Libo~Fu\affil{3}, Lin~Wu\affil{2}, Li~Deng\affil{2}}

\affiliation{1}{Purple Mountain Observatory, Chinese Academy of Sciences, Nanjing 210023, China}
\affiliation{2}{State Key Laboratory of Space Weather, National Space Science Center, Chinese Academy of Sciences, Beijing 100190, China}
\affiliation{3}{Institute of Space Science and Applied Technology, Harbin Institute of Technology, Shenzhen 518055, China}
\affiliation{4}{Laboratory for Electromagnetic Detection, Institute of Space Sciences, Shandong University, Weihai, Shandong 264209, China}
\affiliation{5}{Yunnan Key Laboratory of the Solar physics and Space Science, Kunming 650216, China}
\affiliation{6}{Yunnan Observatories, Chinese Academy of Sciences, Kunming 650216, China}
\affiliation{7}{School of Earth and Space Sciences, Peking University, Beijing 100871, China}
\affiliation{8}{School of Astronomy and Space Science, Nanjing University, Nanjing 210023, China}
\affiliation{9}{Key Laboratory of Particle Astrophysics, Institute of High Energy Physics, Chinese Academy of Sciences, Beijing 100049, China}

\correspondingauthor{Dong~Li}{lidong@pmo.ac.cn}

\begin{keypoints}
\item Multiple periods are identified in a wide range of radio frequencies, and the used radio telescopes are supported by Chinese Meridian Project.
\item A periodicity at about 3~minutes is simultaneously identified in wavelengths of radio, hard X-ray (HXR), Ultraviolet (UV), and Extreme UV (EUV).
\item A periodicity at about 8~minutes is simultaneously identified in wavelengths of radio, HXR, EUV/UV, Ly$\alpha$ and white light.
\end{keypoints}

\begin{abstract}
We explored the quasi-periodic pulsations (QPPs) at multiple periods
during an X4.0 flare on 2024 May 10 (SOL2024-05-10T06:27), which
occurred in the complex active region of NOAA~13664. The flare
radiation reveals five prominent periods in multiple wavelengths. A
8-min QPP is simultaneously detected in wavelengths of HXR, radio,
UV/EUV, Ly$\alpha$, and white light, which may be associated with
nonthermal electrons periodically accelerated by intermittent
magnetic reconnection that is modulated by the slow wave. A
quasi-period at 14~minutes is observed in the SXR and
high-temperature EUV wavebands, and it may be caused by repeatedly
heated plasmas in hot flare loops. A quasi-period at about
18~minutes is only observed by STIX, with reconstructed SXR images
suggesting that the 18-min period pulsations should be considered as
different flares. Meanwhile, a 3-min QPP is simultaneously detected
in wavelengths of HXR, radio, and UV/EUV, which is directly
modulated by the slow magnetoacoustic wave leaking from sunspot
umbrae. At last, a 2-min QPP is simultaneously detected in HXR and
radio emissions during the pre-flare phase, which is possibly
generated by a quasi-periodic regime of magnetic reconnection that
is triggered by the kink wave.
\end{abstract}

\section*{Plain Language Summary}
Combined various solar telescopes, especially the radio telescopes
supported by Chinese Meridian Project, we analyzed the evolution of
multi-wavelength flaring emission during 06:00$-$07:30 UT on 10 May
2024. Multiple periodicities are identified in the multi-wavelength
light curves. A 2-min period is identified in what is described as a
precursor event in hard X-ray (HXR) and radio wavelengths. A
periodicity at 3~minutes is identified in HXR, radio, Ultraviolet
(UV), and Extreme UV (EUV) wavelength observations. A periodicity at
8~minutes is identified in the HXR, radio, EUV, UV, Lyman alpha and
white light wavelength observations. Further, a 14-min period is
identified in the soft X-ray and high-temperature EUV light curves.
Finally, a 18-min period is identified in the low energy soft
X-rays. We also tried to explore the origin of these periods
based on multi-wavelength observations.

\section{Introduction}
A flare eruption refers to the impulsive and dramatic
electromagnetic radiation over a broad range of wavebands on the Sun
and stars, which could be observed in various heights of the
solar/stellar atmosphere \cite<see>[and reference
therein]{Benz17,Kowalski24}. The flare energy is presumably
transferred from the free magnetic energy that has been stored in
non-potential magnetic configuration in the outer atmosphere
\cite<i.e.,>{Shibata11,Yan22,Lu24}. Generally, the flare energy
releases explosively and suddenly in a short time via the well-known
magnetic reconnection. A portion of the released energy heats
coronal plasmas to a higher temperature of several million Kelvin
(MK) at the timescale of seconds or minutes, and some other energies
rapidly accelerate nonthermal particles to a higher energy in the
range from keV to GeV. The majority of accelerated particles could
propagate along the newly-formed flare loop and permeate into the
lower atmosphere, while a portion of accelerated particles may
escape the solar surface and spread into the interplanetary space
along open magnetic field lines. Such energy-released process can be
well explained by the standard two-dimensional (2D) reconnection
model \cite{Priest02,Lin03}. In the spatially-resolved observation
of a solar flare, the loop-like features at high temperatures can be
clearly seen in wavelengths of Extreme Ultraviolet (EUV) and soft
X-rays (SXR). These hot loops are commonly along the closed magnetic
field lines and root in the opposite magnetic fields. Two footpoints
that are connected by the hot loop are often found in wavebands of
hard X-rays (HXR) and microwave, and a loop-top source may appear in
HXR and microwave emission
\cite<i.e.,>{Masuda94,Chen17,Fleishman20,Li21a}. Meanwhile, double
ribbon-like structures can be observed in the wavelength of
ultraviolet (UV), and the flare ribbons often appear as bright
kernels in the H$\alpha$ or white-light emission
\cite{Temmer07,Lit17,Zhang24}. The so-called `double-ribbons flare'
is the most common flare on the Sun, and it matches the standard 2D
reconnection model. The solar flare that is observed in the optical
continuum channel is usually regarded as the white-light flare
(WLF), which is rare event compared to the SXR flare on the Sun,
largely due to the strong visible background in the photosphere
\cite{Song20,Joshi21,Fremstad23}. Therefore, the WLF is often
detected in the powerful flare on the Sun \cite{Zhao21}, or it can
be observed in the superflare in solar-type stars
\cite{Shibayama13,Yan21}.

The feature of quasi-periodic pulsations (QPPs) is a common
phenomenon that is usually associated with the flare radiation in
multiple wavelengths. In a broad sense, the QPPs, which are
time-dependent oscillations, are often defined as a sequence of
repetitive and successive pulsations or pulses in light curves of
solar/stellar flares \cite<cf.>{Zimovets21}. A typical QPP event
should have at least three or four successive pulsations in the time
series. It is not necessary to discuss the periodic behavior when
there are just one or two pulses, which might be just a coincidence
occurred by chance \cite{McLaughlin18,Li24a}. The time interval for
each pulsation of a flare QPP is expected to be roughly equal, which
is termed as the period. However, the majority durations of detected
pulsations are always non-stationary, regarded as the quasi-period
of non-stationarity QPPs \cite{Nakariakov19,Mehta23}. Their
quasi-periods have been reported in a broad range of timescales,
i.e., dozens of milliseconds, tens of seconds, and even a few tens
minutes
\cite<i.e.,>{Tan10,Nakariakov18,Kashapova20,Li21,Li23,Karlicky23,Collier24,Huang24,Inglis24,Li25}.
Similar to the flare radiation, the flare QPPs on the Sun can be
found in a wide range of wavelengths, such as radio/microwave, white
light, H$\alpha$, Ly$\alpha$, UV/EUV, SXR/HXR, and $\gamma$-rays
\cite<i.e.,>{Milligan17,Chelpanov21,Li22a,Li22,Shen22,Zimovets23,Shi24,Millar24,Zhou24}.
On the other hand, the flare QPPs in solar-type stars are often
reported in wavebands of white light and X-rays
\cite{Kolotkov21,Howard22}. The quasi-periods measured in solar and
stellar flares often depend on the time resolution of the detected
telescopes \cite{McLaughlin18}.

It is still an open issue for the generation mechanism of flare QPPs
\cite{Zimovets21,Inglis23}. The QPPs behavior can be directly
induced by the magnetohydrodynamic (MHD) wave in plasma/magnetic
loops, and the eigenmodes may be slow-mode, fast kink- or
sausage-mode \cite{Nakariakov20}. The QPPs feature can also be
driven by a periodic regime of intermittent magnetic reconnection,
and the periodic reconnection might be either triggered or
spontaneous
\cite<i.e.,>{Karampelas23,Comisso24,Kumar24,Sharma24,Zhang24b}. That
is, the periodicity of magnetic reconnection could be caused owing
to an external MHD wave, or it might be regarded as a
self-oscillation system. In such case, the nonthermal electrons and
ions are periodically accelerated by the intermittent magnetic
reconnection during the solar flare. So, it is easier to trigger the
QPPs feature that is observed in HXR and microwave channels during
the impulsive phase of a solar flare \cite{Yuan19,Luo22,Li24b}. In a
recent review paper \cite{Zimovets21}, fifteen mechanisms/models are
summarized, which are almost according to the MHD waves or the
intermittent reconnection model. However, it is still impossible to
determine an unambiguous conclusion that which generation mechanism
should be responsible for all flare QPPs, largely due to the
qualitative nature of generation models and the insufficient
information of observational data. That is, it is not yet possible
to use one single model to interpret all observed QPPs, and the
detected QPPs in different categories could be caused by different
mechanisms \cite{McLaughlin18,Inglis23}. Because the available
observations can not be fully distinguish from various models.

The observed QPPs show a broad range of quasi-periods, and one QPP
event is often dominated by a certain period. The QPPs with double
or multiple periods are also reported in a same event
\cite{Chowdhury15,Karlicky20}. The QPPs at double
periods are thought to be dependent on the MHD modes, and they are
expected to have a period ratio of about two in the weakly
dispersive modes \cite{Nakariakov05}. However, the observed period
ratio often deviates obviously from two, which could be attributed
to the highly dispersive modes or some additional physical effects,
such as the expansion of plasma loops \cite{Verth08}, or the
stratification of longitudinal density \cite{Andries05}. The flare
QPPs with multi-periods could be observed in the same phase or
different phases \cite{Hayes16,Tian16}, and multiple periods in a
same flare could be detected either in one certain wavelength or in
various wavebands \cite{Li17m,Chen19}. Besides these discrete
multi-periods, the continuous multi-periods are also discovered in
flare QPPs, which often reveal a growing periodicity with time, and
the growing periods could be associated with the length of flare
loops \cite{Reznikova11,Pugh19,Li21}. The idea is that the
reconnection site moves toward higher altitudes, therefore, the
flare loops become longer and their footpoints, i.e., the flare
ribbons, moving away from each other.

One critical issue is to understand the generation mechanism of
flare QPPs at multiple periods. In this article, we explored the
QPPs feature with multiple periods during a powerful solar flare, we
also tried to explore their generation mechanisms. The article is
organized as follows: Section~2 describes the observations,
Section~3 introduces the data reduction and our main results,
Section~4 provides some discussions, and a brief summary is shown in
Section~5.

\section{Observations}
On 2024 May 10, a powerful solar flare occurred in the active region
of NOAA~13664, which was a super active region and produced a number
of major flares \cite{Liy24}. The target flare occurred near a group
of sunspots, and it was simultaneously observed by various
instruments: the Atmospheric Imaging Assembly \cite<AIA;>{Lemen12},
the Helioseismic and Magnetic Imager \cite<HMI;>{Schou12}, and the
Extreme ultraviolet Variability Experiment \cite<EVE;>{Woods12} for
the the Solar Dynamics Observatory (SDO), the Solar Ultraviolet
Imager (SUVI) and the X-Ray Sensor (XRS) carried by the
Geostationary Operational Environmental Satellite (GOES), the
Spectrometer/Telescope for Imaging X-rays \cite<STIX;>{Krucker20} on
board the Solar Orbiter, the Gravitational Wave High-Energy
Electromagnetic Counterpart All-sky Monitor \cite<GECAM;>{Xiao22},
the Hard X-ray Imager \cite<HXI;>{Suy19} and the Ly$\alpha$ Solar
Telescope \cite<LST;>{Feng19} for the Advanced Space-based Solar
Observatory \cite<ASO-S;>{Gan19,Huang19}, the New Vacuum Solar
Telescope \cite<NVST;>{Liu14,Yan20}, the Solar Upper Transition
Region Imager \cite<SUTRI;>{Bai23}, the Chinese H$\alpha$ Solar
Explorer \cite<CHASE;>{Lic19}, the Chashan Broadband Solar radio
spectrometer \cite<CBS;>{Shang22,Yan23}, the STEREO/WAVES
\cite<SWAVES;>{Kaiser08}, the Large-Yield RAdiometer (LYRA) on board
the PRoject for OnBoard Autonomy~2 \cite{Dominique13}, the Nobeyama
Radio Polarimeters (NoRP), and the DAocheng Solar Radio Telescope
\cite<DART;>{Yan23b,Lir24}.

SDO/AIA captures EUV/UV snapshots nearly simultaneously with
a temporal cadence of 12/24~s in EUV/UV wavebands, SDO/HMI provides
the line-of-sight (LOS) magnetogram and visible continuum maps near
the Fe~6173~{\AA} line at a temporal cadence of 45~s, and they have
a same spatial scale of 0.6$^{\prime\prime}$~pixel$^{-1}$ after
processing by a standard procedure. SDO/EVE records the EUV spectrum
for the entire Sun in the wavelength range of about 1$-$1216~{\AA}.
Some emission lines at EUV and Ly$\alpha$ can be extracted from the
EUV spectrum, and their temporal cadence is averaged down to 60~s.
GOES/SUVI probes the corona that includes million-degree
temperature, which has a temporal cadence of $\sim$120~s, and a
spatial scale of 2.5$^{\prime\prime}$~pixel$^{-1}$. GOES/XRS records
the SXR fluxes integrated over the whole Sun in 1$-$8~{\AA} and
0.5$-$4~{\AA} at a temporal cadence of 1~s. Based on their ratio,
the iso-thermal temperature could be estimated \cite{White05}. LYRA
records the solar irradiance at four wavebands at a temporal cadence
of 0.05~s. STIX provides the flare imaging spectroscopy of solar
flares in the X-ray range of 4$-$150~keV at a temporal cadence of
about 0.5~s. GECAM can measure solar irradiance in X- and
$\gamma$-rays. In this case, the solar X-ray flux at about
18$-$48~keV was used, which has a temporal cadence of 1~s.

ASO-S/LST equipped with two telescopes: the Solar Disk
Imager (SDI) and the White-light Solar Telescope (WST). SDI takes
the Ly$\alpha$ maps with a normal cadence of 60~s, and it changes to
$\sim$6~s in flare mode. WST provides the white-light map at
3600~{\AA} in a normal cadence of 120~s, and it can reach to
$\sim$1~s in the flare mode. HXI provides the imaging spectroscopy
of solar flares in HXR channels of about 10-300~keV. The temporal
cadence is normally 4~s, and it can be as high as 0.125~s in the
flare mode. NVST, which situates at Fuxian Lake, can provide the
localized images in channels of H$\alpha$ and TiO-band. In this
case, the TiO images with a FOV of
133$^{\prime\prime}$$\times$112$^{\prime\prime}$ were provided, and
it has a spatial scale of 0.052$^{\prime\prime}$~pixel$^{-1}$ and a
temporal cadence of $\sim$30~s. CHASE operates the spectroscopic
observation of the entire Sun in wavelength ranges of H$\alpha$ and
Fe I at a temporal cadence of about 71~s and a spatial scale of
1.04$^{\prime\prime}$~pixel$^{-1}$. SUTRI takes the EUV map at a
temperature of $\sim$0.5 MK \cite{Tian17}, which has a temporal
cadence of $\sim$31~s and a spatial scale of
1.23$^{\prime\prime}$~pixel$^{-1}$. Noting that the CHASE and SUTRI
maps are discontinuous due to the orbit period, and they missed the
impulsive phase of the major flare.

The DART Radiaheliograph routinely produces solar radio
snapshots from 149~MHz to 447~MHz. The spatial scale is about
52$^{\prime\prime}$~pixel$^{-1}$ in the frequency of 447~MHz, and it
decreases to about 157$^{\prime\prime}$~pixel$^{-1}$ at 149~MHz,
while it has a uniform temporal cadence of 10~s. CBS is a newly
built solar radio spectrograph at Chashan Solar radio Observatory
(CSO). It provides solar radio dynamic spectra in a broad frequency
range of millimeter ($\sim$35$-$40~GHz, mm), centimeter
($\sim$6$-$15~GHz, cm), decimeter ($\sim$0.5$-$6~GHz, dm),
meter-decameter ($\sim$80$-$610~MHz, m) regimes. Their temporal
cadences are $\sim$0.53~s (CBSmm), $\sim$0.067~s (CBScm),
$\sim$2.15~s (CBSdm), and $\sim$0.1~s (CBSm), respectively. DART and
CSO are both solar radio telescopes supported by Chinese initiative
Meridian Project phase~2. SWAVES takes the solar radio spectrum in
the low frequency range of $\sim$0.0026$-$16.025~MHz at a temporal
cadence of 60~s. NoRP measures solar radio fluxes in six microwave
frequencies at a temporal cadence of 1~s.

\section{Data reduction and results}
The target flare was measured by various instruments, which provided
us an opportunity to investigate the flare QPPs at multiple periods
in multiple wavelengths, i.e., HXR/SXR, EUV/UV, white light,
Ly$\alpha$, and radio.

\subsection{Overview of the powerful flare}
\begin{figure}
\centering
\includegraphics[width=\textwidth]{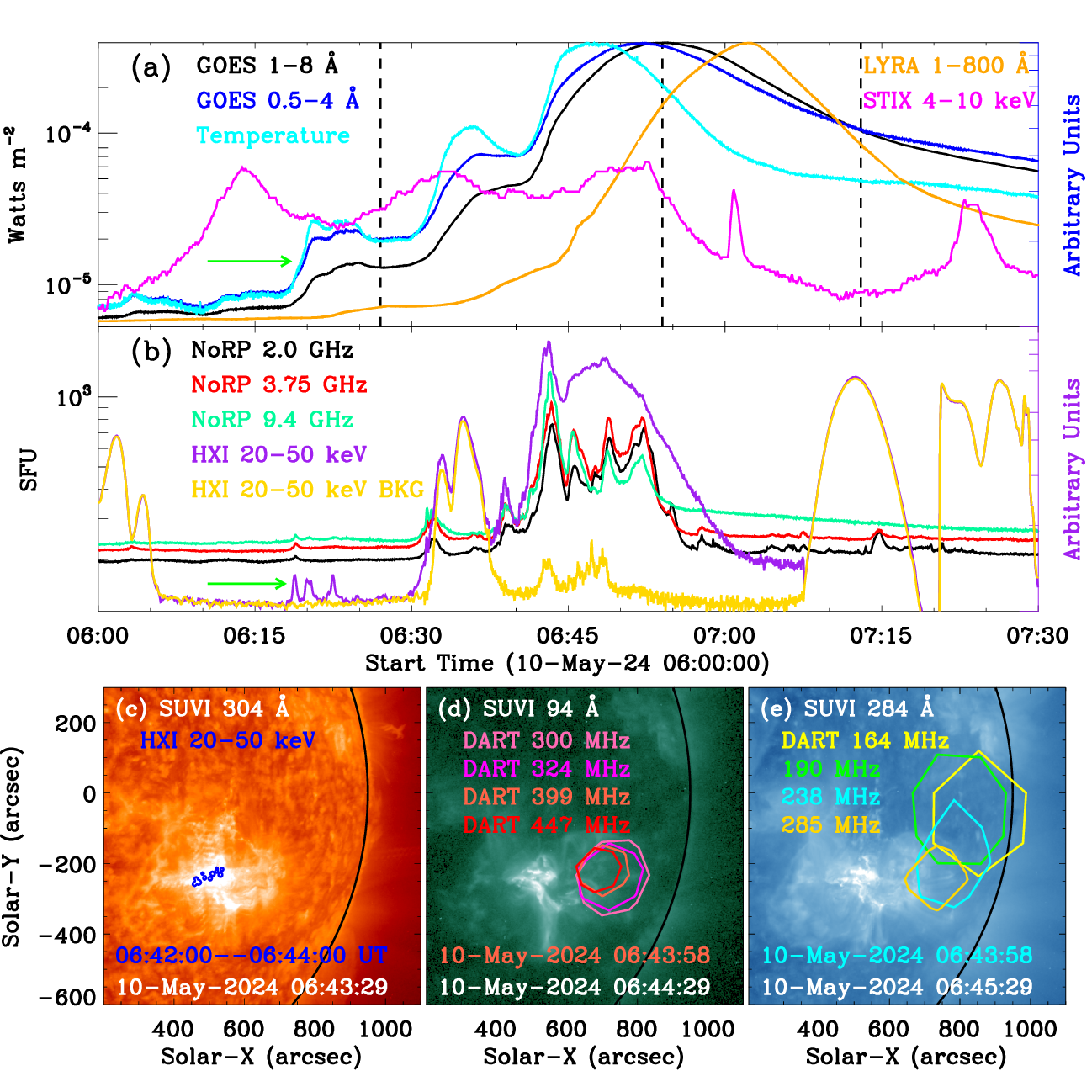}
\caption{Overview of a powerful flare occurred on 2024 May 10. (a):
Light curves measured by GOES~1-8~{\AA}, 0.5-4~{\AA}, and
temperature, LYRA~1-800~{\AA}, and STIX~4-10~keV. The vertical lines
mark the start, peak and stop times of the powerful flare. (b): Time
series recorded by NoRP~2.0~GHz, 3.75~GHz, and 9.4~GHz,
HXI~20-50~keV and its background. The green arrow indicates a flare
precursor. (c)-(e): EUV snapshots with a large FOV of
900$^{\prime\prime}$$\times$900$^{\prime\prime}$ measured by
GOES/SUVI in wavelengths of 304~{\AA}, 94~{\AA}, and 284~{\AA}. The
blue contours represent the flare radiation in the HXR channel, and
the contour level is set at 10\%. The colored contours in panels~(d)
and (e) outlines the radio radiation observed by DART, and their
contour levels are set at 60\%. \label{over}}
\end{figure}

Figure~\ref{over} presents an overview of the target flare on 2024
May 10. Panel~(a) shows the full-disk light curves in SXR and X-ray
ultraviolet (XUV) wavebands from 06:00 to 07:30~UT. The
GOES~1$-$8~{\AA} flux suggests an X4.0-class flare, it started at
about 06:27~UT, reached its maximum at about 06:54~UT, and stopped
at about 07:13~UT, as marked by the vertical dashed lines.
The GOES fluxes at 1$-$8~{\AA} and 0.5$-$4~{\AA} reveals
three successive pulsations from about 06:18 to 07:00~UT, indicating
a QPP signal in the SXR channel, and it is much more visible in the
temperature profile, suggesting that it highly depends on periodic
variations of the plasma temperature. The STIX flux at 4-10~keV also
shows three successive pulsations during $\sim$06:01$-$06:59~UT.
Obviously, the QPP period is longer than that observed by GOES.
Conversely, XUV flux at 1$-$800~{\AA} recorded by LYRA appears a
rather weak signal of QPP.

In Figure~\ref{over}(b), we show the light curves in channels of HXR
and microwave measured by HXI and NoRP, which also appear a number
of successive pulsations. However, they are different from those
successive pulsations in the SXR channel. Moreover, at least three
successive pulsations with a small amplitude are seen in the HXR
channel, as indicated by the green arrow. They appear before the
onset of the X4.0 flare, which might be regarded as a precursor.
Panels~(c)$-$(e) show EUV maps observed by GOES/SUVI during the
flare impulsive phase, and they have a large FOV. The over-plotted
blue contours represent the HXR radiation, which are reconstructed
by the HXI\_CLEAN algorithm from the HXI data between
06:42$-$06:44~UT. The flare sources seen in wavebands of HXR and EUV
match with each other. We also plot the radio sources in the
meter-wave regime measured by DART, as shown by the color contours
in panels~(d) and (e). The radio sources are far away from the flare
area. The distance between the radio source and the flare site
becomes farther with the observed frequency decreasing, but the
radio source region becomes larger and larger. This feature could be
attributed to the angular resolution of DART, that is, the angular
resolution of DART becomes lower and lower with the decreasing of
radio frequencies. Or it may be due to the high loop top
\cite{Gary18}.

\begin{figure}
\centering
\includegraphics[width=\textwidth]{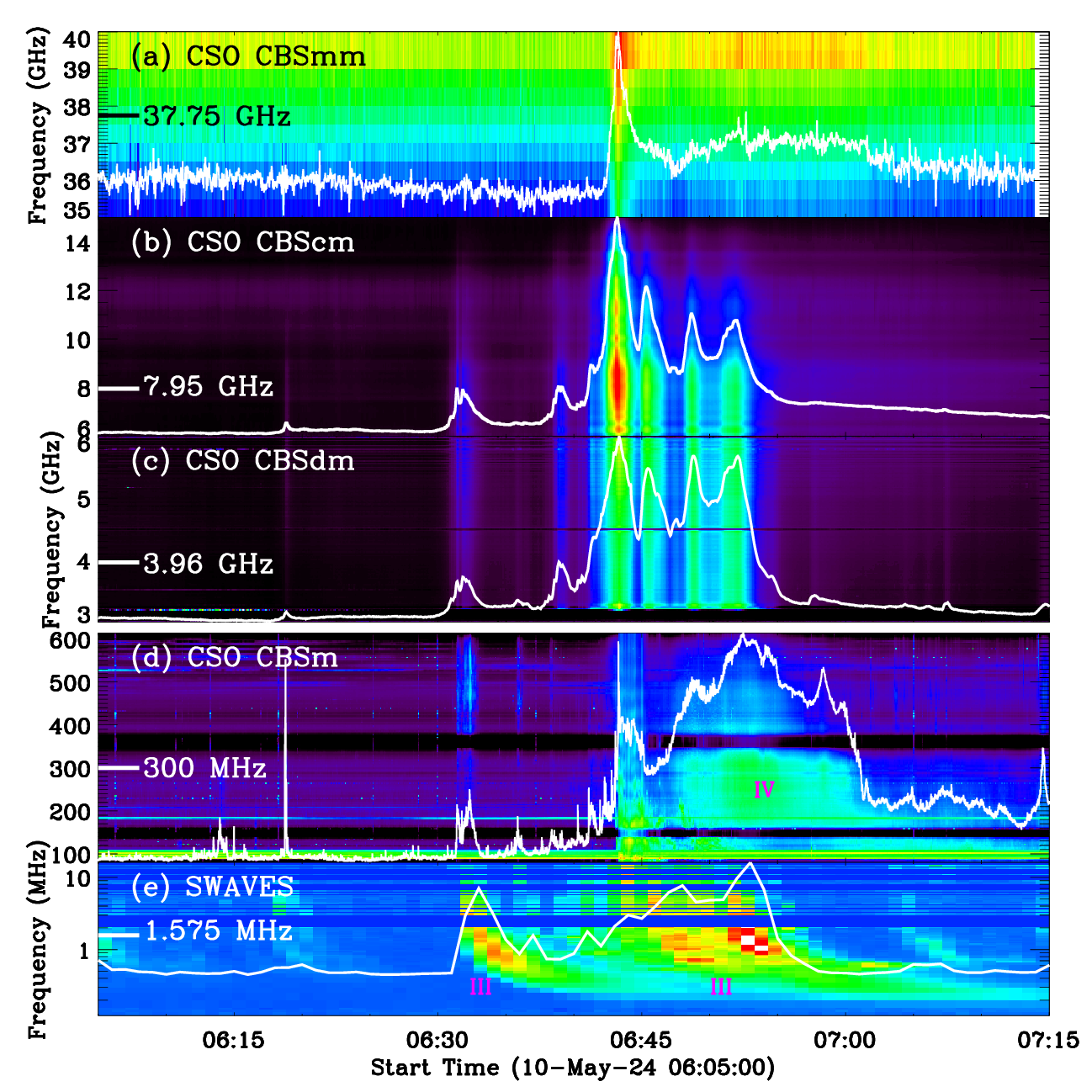}
\caption{Dynamic radio spectra measured by CSO Radiaheliograph
(millimeter, centimeter, decimeter, and meter-decameter regimes) and
SWAVES. The overplotted curves are the radio fluxes derived from the
radio spectra, as marked by the short line in each panel.
\label{spec}}
\end{figure}

Figure~\ref{spec} shows the dynamic radio spectra measured by CSO
Radiaheliograph and SWAVES in a wide range of frequencies. The
over-plotted curve in each panel represents the radio flux at the
designated frequency, as marked by the short line on the left hand.
A sequence of radio bursts can be seen in the dynamic spectra, and
they drift rapidly from higher to lower frequencies, which might be
regarded as a group of centimeter-decimeter bursts, as shown in
panels~(a)$-$(e). On the other hand, some radio bursts only appear
in the meter-decameter regime (panel~d), and they are almost no
frequency drift, which might be considered as the type IV burst. A
type III group is seen in the low frequency range measured by
SWAVES, as shown in panel~(e). Anyway, all these radio emissions
seem to show quasi-periodicity. The overlaid fluxes show a number of
successive pulsations, especially for the radio fluxes in
frequencies of 7.95~GHz, 3.96~GHz, 300~MHz, and 1.575~MHz.
Conversely, the radio flux at 37.75~GHz show a weak signal of QPP,
which may be attributed to its higher frequency, and thus it
requires a large amount of energies.

\subsection{Multi-periodic pulsations}
\begin{figure}
\centering
\includegraphics[width=\textwidth]{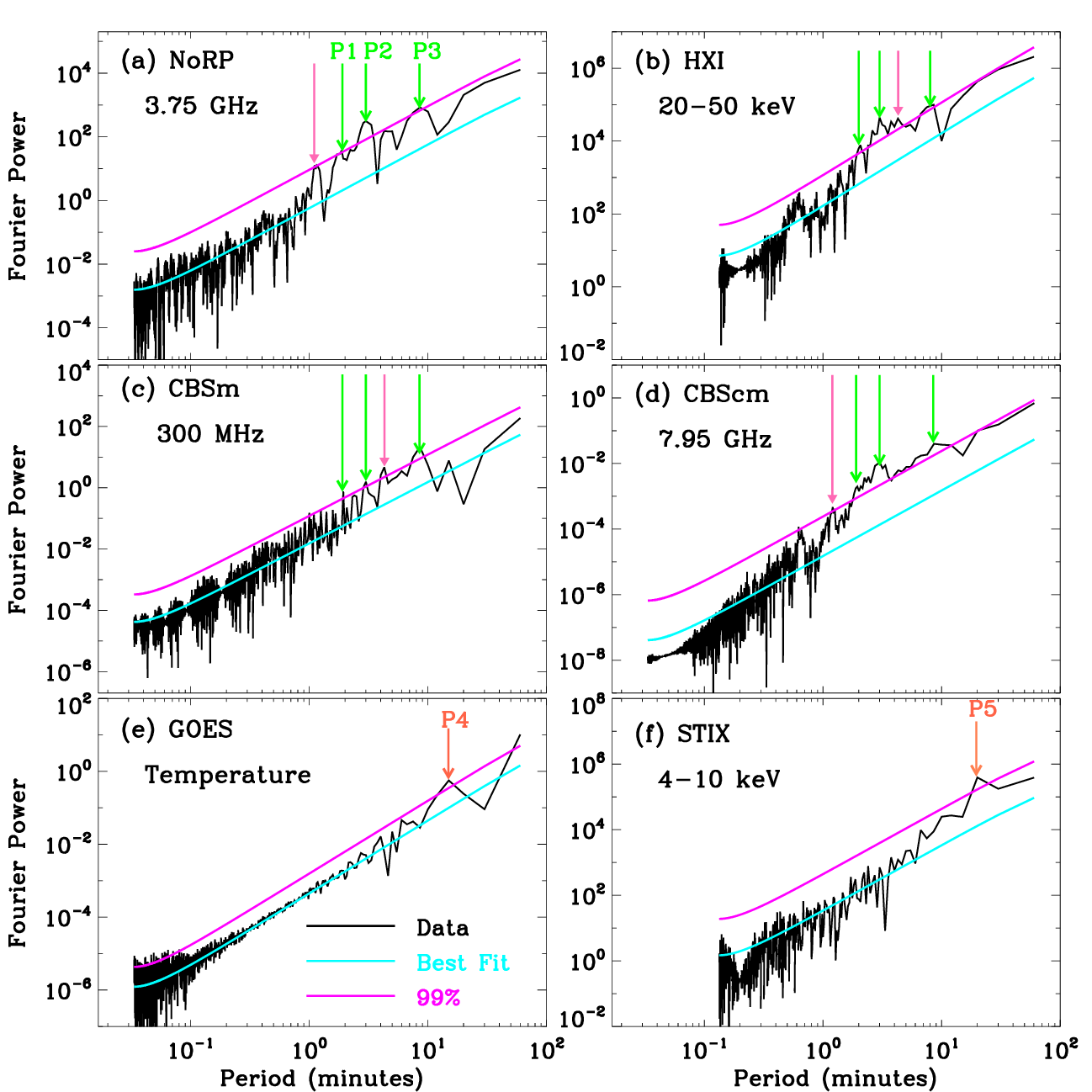}
\caption{Fourier power spectra of raw light curves in multiple
wavelengths. The cyan line in each panel represents a best-fit
result for the observational data (black), and the magenta line
indicates the confidence level at 99\%. The color arrows mark
quasi-periods for P1-P5 above the confidence level. \label{fft1}}
\end{figure}

In order to determine the quasi-period of flare QPPs, we perform the
fast Fourier Transform (FFT) for the raw light curves with a
Lomb-Scargle periodogram method \cite{Scargle82}, and the confidence
level is defined by \citeA{Horne86}. Figure~\ref{fft1} presents the
FFT power spectra in multiple wavebands, the cyan curve represents
the best fit result for the observational data, and the magenta
curve is the confidence level at 99\%. We can find that several
peaks exceed the 99\% confidence level, especially in wavebands of
radio and HXR (a-d), as indicated by the color arrows. However, some
peaks only appear in one or two wavebands (hot pink arrows), so they
are not considered in this study. At last, three dominant periods
(P1-P3) are simultaneously identified in NoRP~3.75~GHz,
HXI~20-50~keV, CBS~300~MHz and CBS~7.95~GHz, as indicated by the
green arrows. For simple, the three dominant periods are regarded as
2-, 3-, and 8-minutes, respectively.

In Figure~\ref{fft1}~(e) and (f), we also show the FFT power spectra
in the SXR channel. From which, a dominant period of about
14~minutes (P4) appears in the time series of GOES temperature, and
a dominant period of about 18~minutes (P5) exceeds the confidence
level in the energy range of STIX~4-10~keV. The FFT results are
consistent with the three pulsations in raw light curves measured by
GOES and STIX in Figure~\ref{over}, confirming the existence of
14-min and 18-min QPPs.

\subsection{Local observations}
\begin{figure}
\centering
\includegraphics[width=\textwidth]{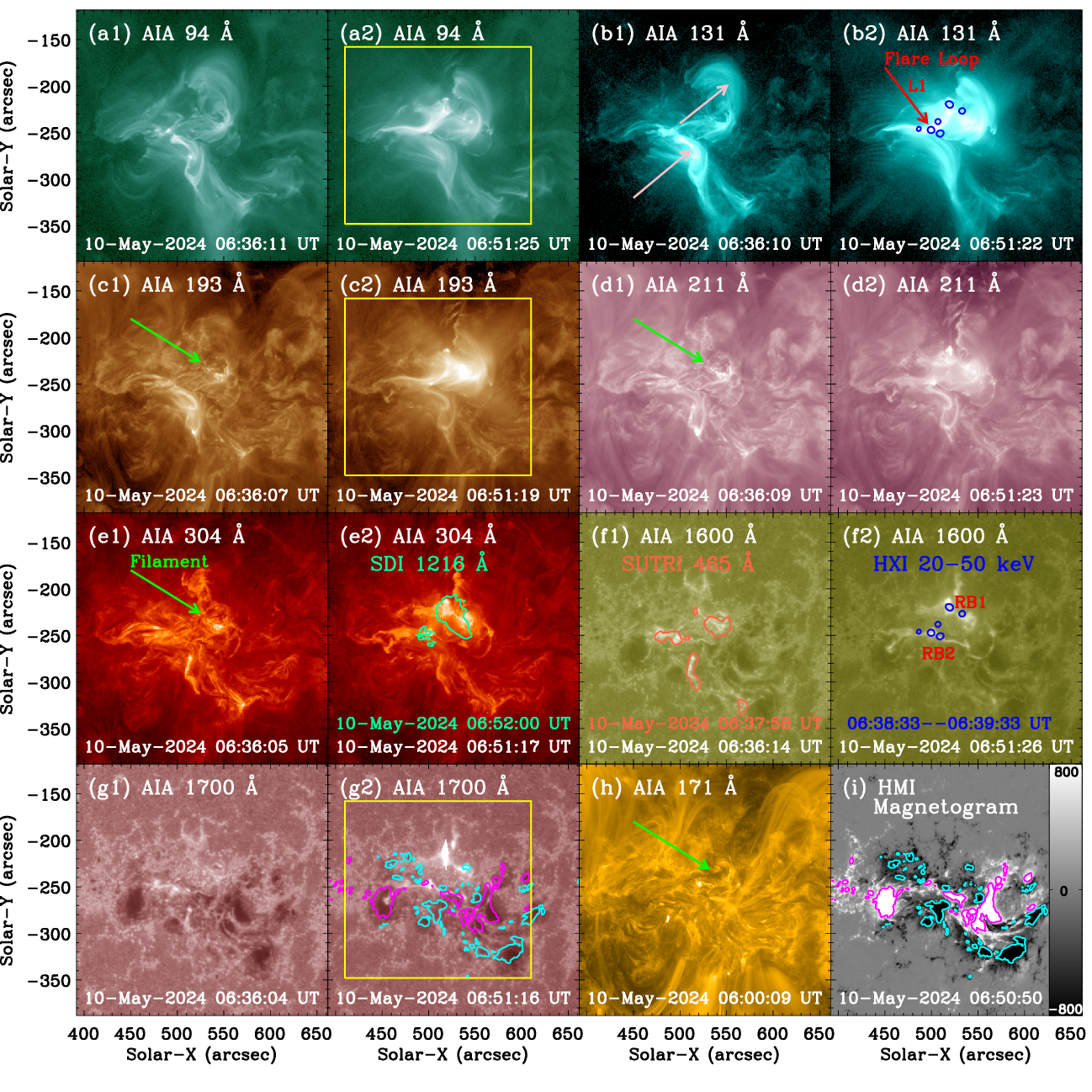}
\caption{(a1)-(g2): EUV/UV snapshots captured by SDO/AIA at
$\sim$06:36~UT and $\sim$06:51~UT during the X4.0 flare. (h): The
AIA~171~{\AA} map at about 06:00~UT before the X4.0 flare. (i): The
LOS magnetogram measured by SDO/HMI. They have a same FOV of about
270$^{\prime\prime}$$\times$270$^{\prime\prime}$. The blue contours
represent the HXR emission at HXI~20-50~keV, and the level is set at
50\%. The magenta and cyan contours are the positive and negative
magnetic fields at a strength of $\pm$800~G. The pink and red arrow
marks the hot flare loop, and the green arrow indicates a filament.
The yellow rectangle outlines the flare area used to integrate the
intensity curve observed by SDO/AIA and ASO-S/SDI. \label{img1}}
\end{figure}

\begin{figure}
\centering
\includegraphics[width=\textwidth]{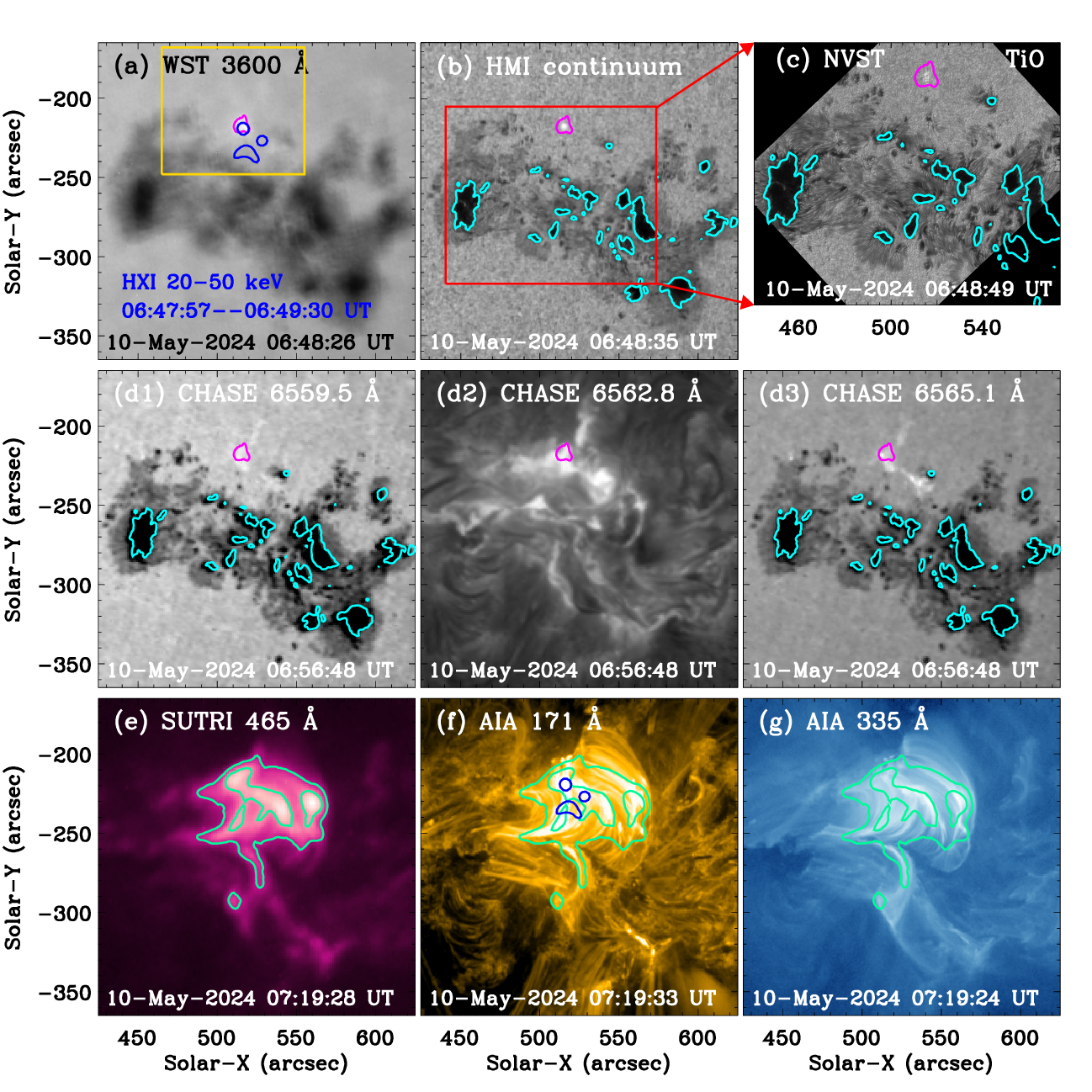}
\caption{Multi-wavelength snapshots measured by ASO-S/WST, SDO/HMI,
NVST, CHASE, SUTRI, and SDO/AIA. They have a small FOV of about
200$^{\prime\prime}$$\times$200$^{\prime\prime}$, and the TiO-band
map has a much smaller FOV of
133$^{\prime\prime}$$\times$112$^{\prime\prime}$ due to the
observational limit, as outlined by the red rectangle in panel~(b).
The magenta contour outlines the white-light brightening in the
wavelength of WST~3600~{\AA}, and the blue contours represent HXR
emissions in the channel of HXI~20-50~keV at levels of 50\%. The
cyan contours outline the sunspot umbra, and the spring green
contours outline the post flare loops. The gold rectangle outlines
the flare area used to integrate the white-light flux. \label{img2}}
\end{figure}

The radio dynamic spectra and HXR/SXR fluxes are all integrated over
the whole Sun. To search for the source region that generated the
flare QPPs, we draw the multi-wavelength images in
Figures~\ref{img1} and \ref{img2}. Figure~\ref{img1} shows the
EUV/UV images and the line-of-sight (LOS) magnetogram with a same
FOV of about 270$^{\prime\prime}$$\times$270$^{\prime\prime}$ during
the X4.0 flare. Two groups of hot loops (pink arrows) can be clearly
seen in wavebands of AIA~94~{\AA} and 131~{\AA} at about 06:36~UT,
and they become a bright loop and several weak loops at about
06:51~UT, as marked by the red arrow. The hot loop can be seen in
AIA~193~{\AA}, but it is hard to be observed in AIA~211~{\AA},
suggesting that the flare loop is filled with high-temperature
plasma, i.e., $>$6~MK. Some bright kernels appear in AIA~304, 1600
and 1700~{\AA}, and SUTRI~465~{\AA} at about 06:36~UT, and they
become double ribbons (RB1 and RB2) at about 06:51~UT. The HXR
emissions observed by HXI~20-50~keV are overlaid on AIA~131 and
1600~{\AA} maps, and they are mainly situated in double ribbons,
named as footpoints, as shown by the blue contours. The
footpoints/ribbons connected by hot loops have changed from about
06:36~UT to 06:51~UT, implying that the magnetic reconnection occurs
during the X4.0 flare. A weak signature of dark structure is
simultaneously found in wavebands of AIA~193, 211, and 304~{\AA} at
about 06:36~UT, as indicated by the green arrow. The dark structure
can be clearly seen in wavebands of AIA~171~{\AA} before the X4.0
flare, i.e., at about 06:00~UT. Moreover, a sunspot group can be
found in AIA~1600 nd 1700~{\AA}, which locates in a complex magnetic
field structures (panel~i). All these observations suggest
that the dark structure could be regarded as a micro-filament, and
the magnetic reconnection during the X4.0 flare could be triggered
by the filament eruption. The gold box outlines the flare area used
to integrate over the local flux in EUV/UV wavelengths.

Figure~\ref{img2} shows some more sub-maps in multiple wavelengths
during the X4.0 flare. In panels~(a)-(c), one patch of visible
continuum enhancement is simultaneously detected in wavebands of
WST~3600~{\AA}, HMI continuum near 6173~{\AA}, and NVST
TiO~7058~{\AA}, as outlined by the magenta contour. Thus, the X4.0
flare can be regarded as a WLF, and the gold rectangle outlines
the flare area used to integrate the white-light flux. Moreover, the
white-light brightening is temporally and spatially consistent with
the HXR radiation source, as marked by the blue contours. The cyan
contours outline the boundary between the umbra and penumbra.
Panels~(d1)-(d3) shows the H$\alpha$ images at its line center and
two line wings measured by CHASE at about 06:56~UT. The radiation
enhancements seen in two line wings of H$\alpha$ (d1 and d3) appear
to match the white-light brightening, suggesting that the H$\alpha$
line-wing radiation is mostly from the upper photosphere to the lower
chromosphere. Similar to the AIA~304~{\AA} image, the H$\alpha$
image at its line center (d2) reveals double ribbons.
Figures~\ref{img2}~(e)-(g) presents the EUV sub-maps observed by
SUTRI~465~{\AA}, AIA~171 and 335~{\AA} at about 07:19~UT. They both
show a number of loop-like structures, which could be regarded as
the post flare loops.

\begin{figure}
\centering
\includegraphics[width=\textwidth]{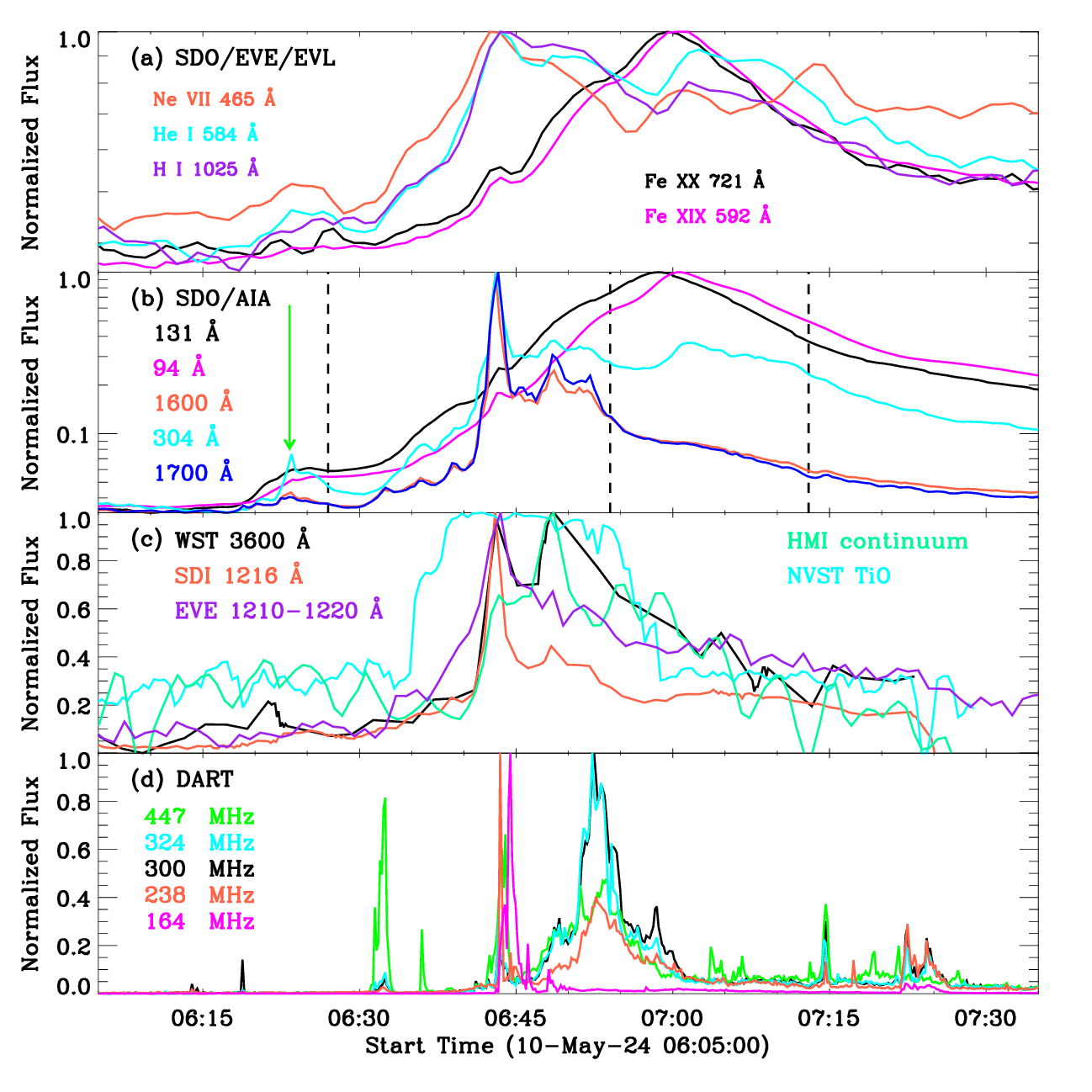}
\caption{Multi-wavelength light curves. (a): Full-disk light curves
recorded by SDO/EVE for the isolated lines. (b): Local intensity
curves integrated over the flare area measured by SDO/AIA. The green
arrow indicates the flare precursor. (c): Local time series
integrated over the flare region observed by WST, SDI, HMI, and
NVST, and the full-disk Ly$\alpha$ flux recorded by SDO/EVE. (d)
Local light curves in radio emissions integrated over the active
region measured by DART. \label{flux}}
\end{figure}

In Figure~\ref{flux}, we show multi-wavelength light curves, and
they are normalized by their maximum intensity. Panel~(a) presents
the full-disk light curves recorded by SDO/EVE for the isolated
lines, including high- (Fe XX, Fe XIX) and low-temperature (Ne~VII,
He~I and H~I) lines. Panel~(b) shows the local light curves measured
by SDO/AIA, which are integrated over the flare area. The local AIA
fluxes are also divided into high- (131 and 94~{\AA}) and
low-temperature (1600, 304 and 1700~{\AA}). The time series at
high-temperature channels show the quasi-period that is consistent
with the SXR fluxes recorded by GOES. Conversely, the time series at
low-temperature channels appear to reveal multi-period QPPs, which
agree with the HXR and radio fluxes. Moreover, the precursor pulse
can also be seen in the local AIA fluxes, as marked by the green
arrow, confirming the presence of flare precursor.
Figure~\ref{flux}~(c) shows the white-light fluxes measured by
WST~3600~{\AA}, HMI continuum, and NVST~TiO, which are integrated
over the WLF region. Here, we also show the local flux in
SDI~1216~{\AA}, and the full-disk light curve in
EVE~1210-1220~{\AA}. They match with each other, regarded as the
Ly$\alpha$ flux. All these time series in wavebands of white light
and Ly$\alpha$ also reveal a prominent signature of
quasi-periodicity. Figure~\ref{flux}~(d) presents the local radio
fluxes at meter-wave regime, which are measured by DART. They all
appear the QPP feature with multiple periods. All these
observational results are similar to that obtained from the FFT
power spectra in Figure~\ref{fft1}.

\subsection{Precursor QPP}
\begin{figure}
\centering
\includegraphics[width=\textwidth]{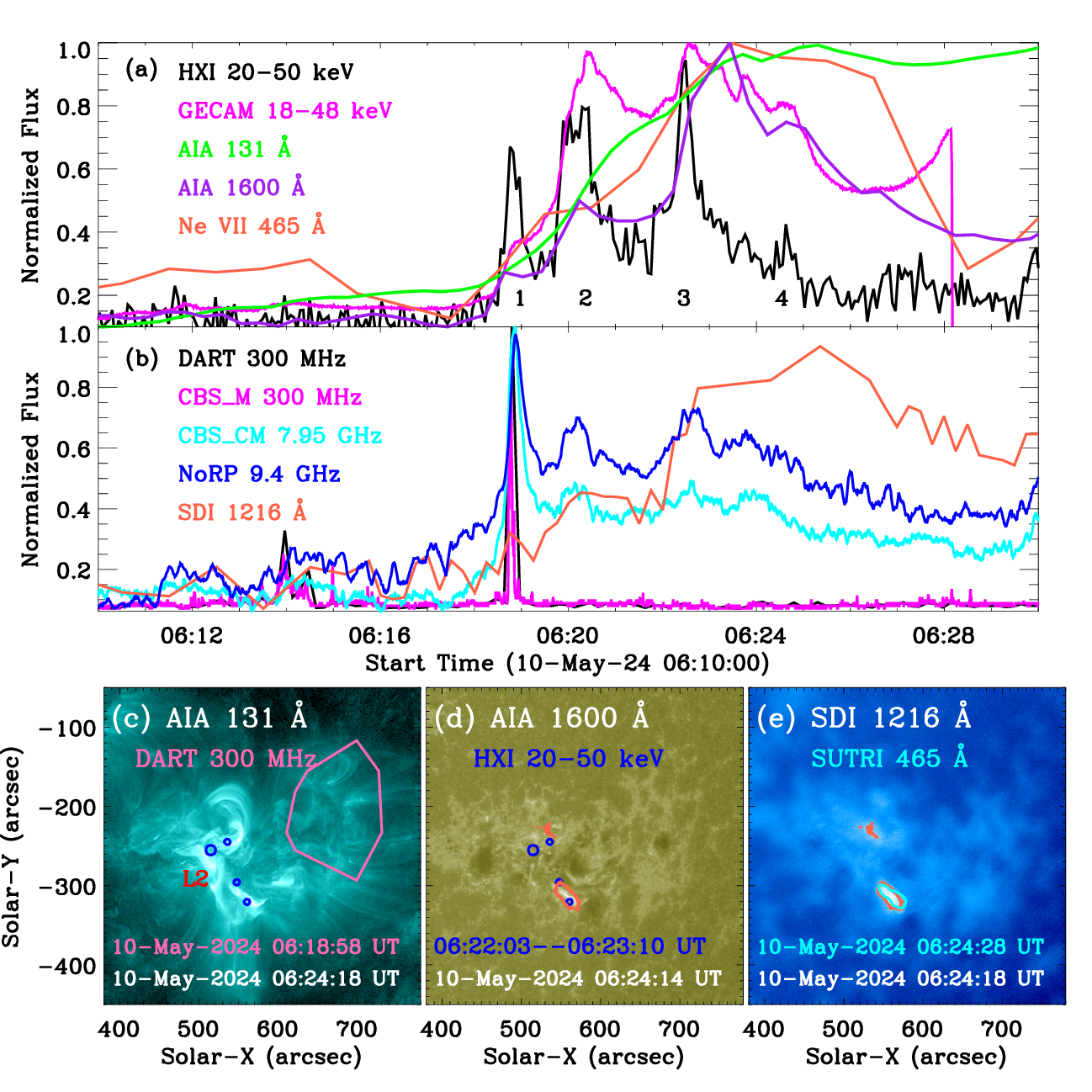}
\caption{Overview of the pre-flare phase. (a) and (b):
Multi-wavelength light curves measured by HXI, GECAM, SDO/AIA,
SDO/EVE, DART, CBS, NoRP, and SDI. (c): Multi-wavelength maps
observed by SDO/AIA and LST/SDI. The overlaid contours outline the
flare radiation measured by HXI~20-50~keV (blue, 10\% level),
DART~300~MHz (hot pink, 80\% level), SDI~1216~{\AA} (tomato), and
SUTRI~465~{\AA} (cyan). \label{pre}}
\end{figure}

The local fluxes measured by SDO and ASO-S demonstrated the
existence of the flare precursor. Figure~\ref{pre} presents the
light curves (a and b) and images (c-e) in multiple wavelengths
during the pre-flare phase, i.e., from 06:10~UT to 06:30~UT. The HXR
light curves in the energy range of HXI~20-50~keV and GECAM
18-48~keV show four successive peaks (1-4) during
$\sim$06:17-06:25~UT, within an average duration of about 2~minutes.
The average duration is consistent with the 2-min period (P1)
detected in radio and HXR channels, which may be regarded as the
precursor QPP. The precursor QPP can be seen in radio/microwave
fluxes in frequencies of NoRP~9.4~GHz and CBS~7.95~GHz, and it shows
a weak QPP signature in the local light curves in wavelengths of
AIA~1600~{\AA} and SDI~1216~{\AA}. However, it is hard to see any
QPP feature in channels of AIA~131~{\AA}, Ne VII~465~{\AA} (EUV),
DART~300~MHz, and CBS~300~MHz (meter-wave). These intensity curves
only show one apparent pulse during the precursor, and the pulse
seen in the meter-wave regime is shorter/narrower and earlier than
that observed in the EUV wavelengths. This observational fact also
implies that the flare precursor is highly related to the nonthermal
electrons, and the precursor QPP could be driven by a quasi-periodic
regime of magnetic reconnection. Panels~(c)-(e) show the EUV/UV
sub-maps with a same FOV of about
400$^{\prime\prime}$$\times$400$^{\prime\prime}$ during the flare
precursor. Similarly to what has seen in the X4.0 flare, some hot
loop-like features can be found in the high-temperature channel of
AIA~131~{\AA}, and some ribbon-like structures are seen in the
low-temperature channel of AIA~1600~{\AA}, SDI~1216~{\AA} (tomato),
and SUTRI~465~{\AA} (cyan). Moreover, some bright kernels (blue) are
measured by HXI~20-50~keV, and they appear to be connected by hot
loops, termed as footpoints. At last, the radio source measured by
DART~300~MHz seems to be far way from the precursor region, which
may be attributed to the high loop top.

\section{Discussions}
The X4.0 flare shows QPP behaviors at multiple periods in various
wavebands during different flare phases. In order to reveal their
trigger mechanisms, we used the wavelet transform and Fourier
transform to determine the temporal and spatial distributions of
multi-period QPPs.

\subsection{Temporal distribution}
\begin{figure}
\centering
\includegraphics[width=\textwidth]{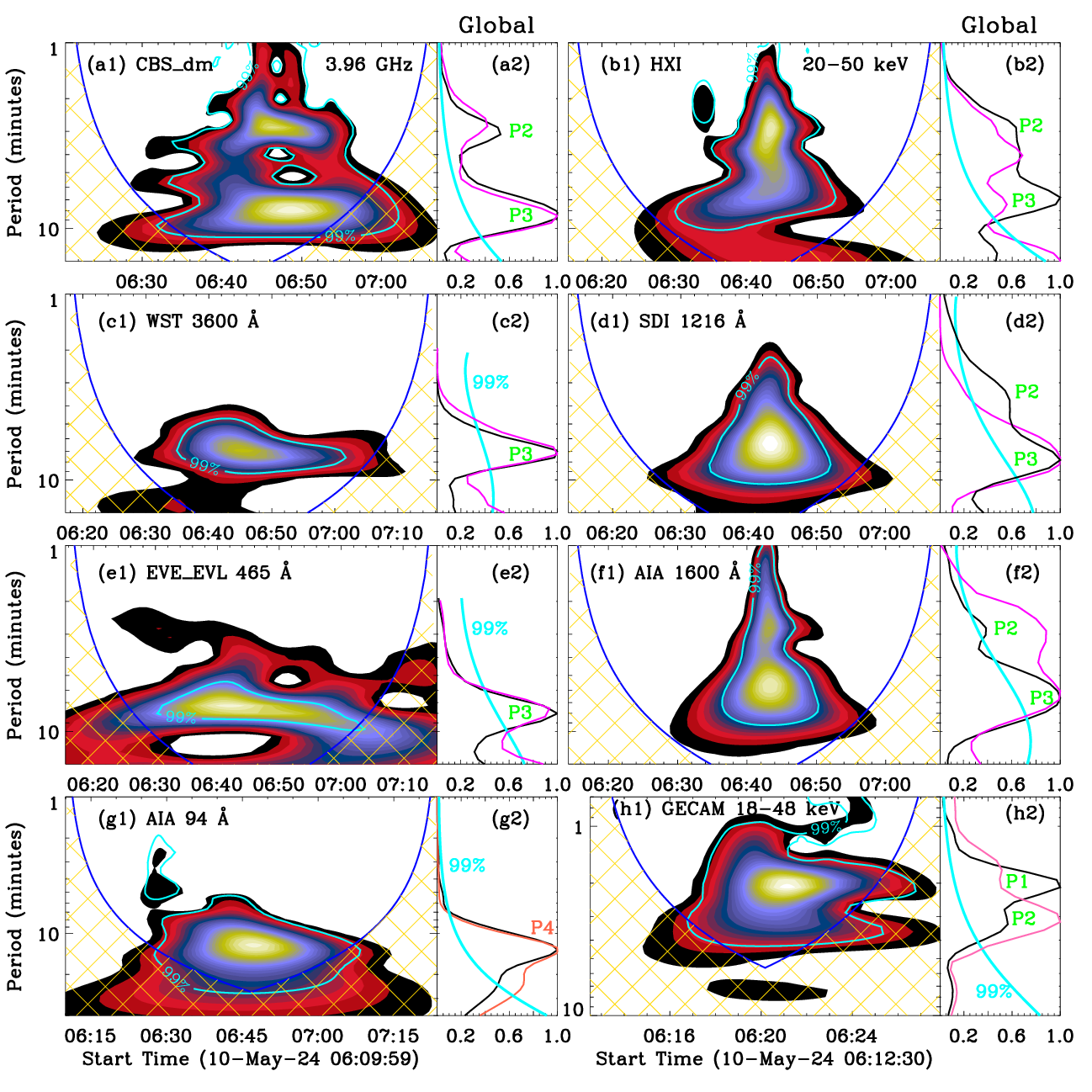}
\caption{Morlet wavelet analysis results. (a1)-(h1): Morlet wavelet
power spectra. (a2)-(f2): Global wavelet power spectra for the
running windows of 10~minutes (black) and 15~minutes (magenta),
respectively. (g2): Global wavelet power spectra for the running
windows of 15~minutes (black) and 20~minutes (tomato). (h2): Global
wavelet power spectra for the running windows of 3~minutes (black)
and 5~minutes (hot pink). The cyan contours and lines represent the
significance level of 99\%. \label{wav}}
\end{figure}

In order to identify the temporal distribution of the flare QPPs, we
performed the wavelet transform with a `mother' function
\cite{Torrence98} for the detrended time series. This method can
well resolve the temporal distribution of periods, that is, the
time-resolved period. However, it is often used for the detrended
time series, which might introduce a spurious periodicity in signals
\cite{Broomhall19}. In this study, two running windows are used for
each raw light curve to obtain the detrended time series, and thus
the artifact period of the detrending process could be excluded
\cite<i.e.,>{Tian12,Li24c}. The time resolution of light curves
measured by WST~3600~{\AA} and SDI~1216~{\AA} is nonuniform due to
the switch of observational modes. Therefore, the white-light and
Ly$\alpha$ fluxes are interpolated into the unform time resolutions
of 120~s and 60~s, respectively. Figure~\ref{wav} shows the Morlet
wavelet analysis results for multi-wavelength light curves.
Considering the broad period range of QPPs, different running
windows are used for various channels. Panels~(a1)-(f1) show the
Morlet wavelet power spectra in wavebands of radio, HXR, white
light, Ly$\alpha$, and low-temperature EUV/UV during the X4.0 flare,
and they are characterized by a bulk of power spectra inside the
99\% significance level. Panels~(a2)-(f2) presents the global
wavelet power spectra with two running windows of 10 (black) and
15~minutes (magenta), respectively. They all reveal a prominent peak
that is centered at about 8~minutes, confirming the presence of the
8-min period (P3). On the other hand, the global wavelet power
spectra measured by CBS~3.96~GHz, HXI~20-50~keV, SDI~1216~{\AA}, and
AIA~1600~{\AA} display another peak that is centered at about
3~minutes, although the signature of 3-min QPP in wavebands of
SDI~1216~{\AA} and AIA~1600~{\AA} is a bit weak. All these results
provide prominent observational signatures of the existence of the
3-min period (P2). On the contrary, the 3-min period is not observed
in wavebands of WST~3600~{\AA} and EVL~465~{\AA}, largely due to
their lower temporal resolution, i.e., 2~minutes.

Panels~(g1) and (g2) present the Morlet wavelet power spectrum and
its global wavelet power spectrum in the wavelength of
high-temperature EUV from the precursor to X4.0 flare. Here, the
running windows of 15 (black) and 20~minutes (tomato) are used, as
we expected a 14-min period. The power spectrum shows a broad
range of period at a center of about 14~minutes, which is similar to
the 14-min period (P4) in SXR channels. Panels~(h1)-(h2) show the
Morlet wavelet power spectrum and its global wavelet power spectrum
in the channel of HXR during the precursor. The running windows of 3
(black) and 5~minutes (hot pink) are used, since the 2-min and
3-min periods are expected. Double peaks that are centered at
2~minutes and 3~minutes can be seen in the global wavelet power
spectrum, similar to the P1 and P2 periods. The
wavelet analysis results suggest that the 2-min period is
mainly from the precursor phase, which might be regarded as a
precursory indicator of the powerful flare
\cite{Tan16,Li20}.

\subsection{Spatial distribution}
\begin{figure}
\centering
\includegraphics[width=\textwidth]{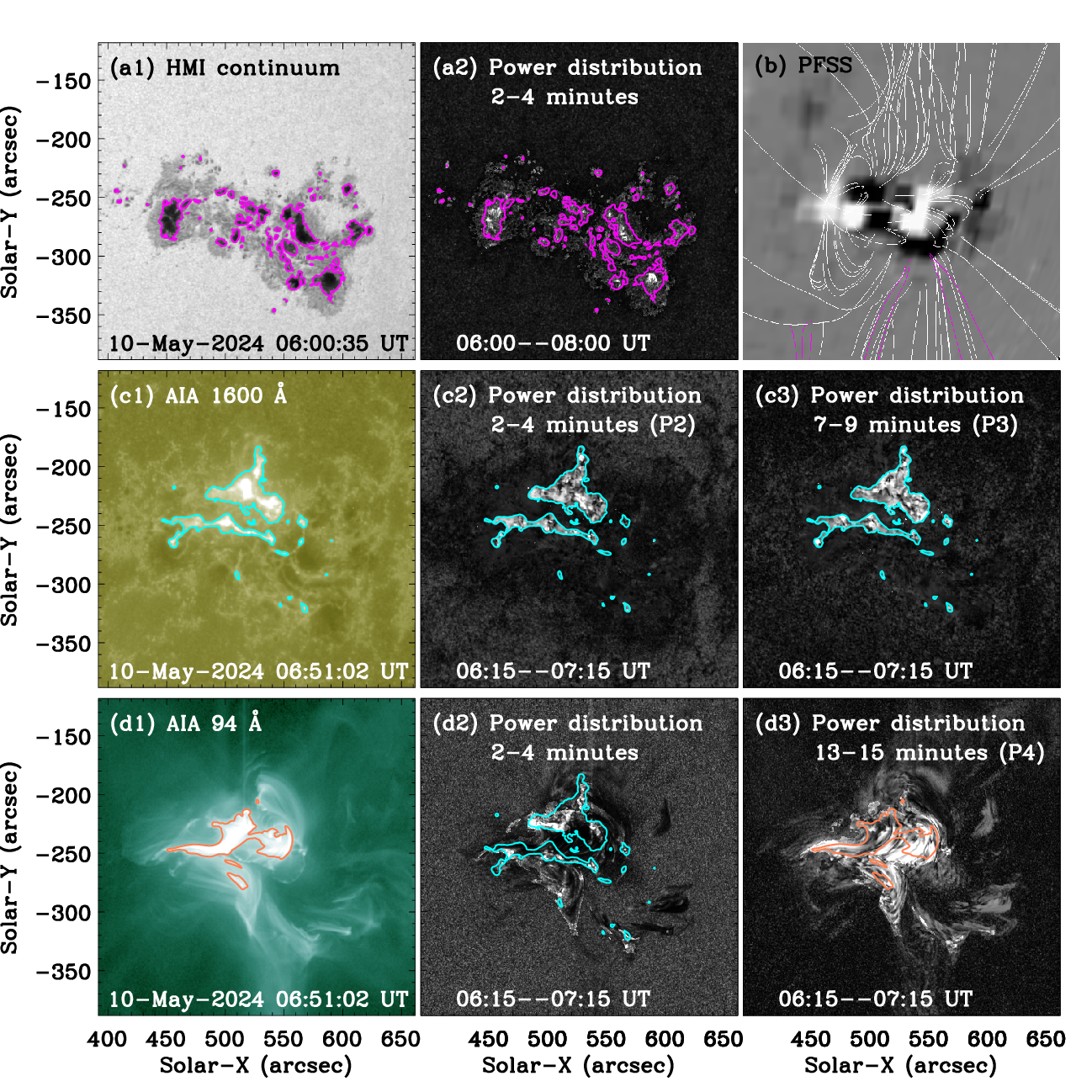}
\caption{(a1)-(d1): Multi-wavelengths images in wavelengths of HMI
continuum, AIA~1600~{\AA}, and 94~{\AA}. The overlaid contours
outline the sunspot umbras (magenta), double flare ribbons (cyan) and
hot flare loops (tomato), respectively. (b): The magnetic field lines
extrapolated from a PFSS model. The purple and white lines indicate
the open and closed magnetic fields, respectively. (a2)-(d3):
Fourier power maps that are averaged over 2-4~minutes (a2-d2) and
7-10~minutes (c3) or 13-15~minutes (d3), respectively. \label{pow}}
\end{figure}

In order to determine the trigger source of flare QPPs at various
periods in multiple wavelengths, that is, spatial-resolved their
sources, the Fourier transform \cite<i.e.,>{Inglis08,Sych08,Feng20}
is performed on the radiation intensity of every pixel in several
channels, as shown in Figure~\ref{pow}. Panel~(a1) shows the visible
continuum map measured by SDO/HMI before the X4.0 flare. It reveals
a group of sunspots, the umbras and penumbras are clearly
distinguished, as indicated by the magenta contours. Panel~(a2)
presents the spatial distribution of the normalized Fourier power
that is averaged over the spectral component at 2-4 minutes for the
HMI continuum data during 06:00-08:00~UT, containing the flare and
non-flare time. We can find that the spectral component at the
period range of 2-4 minutes is generally situated in sunspot umbras,
as outlined by the magenta contours. Panel~(b) presents the magnetic
configuration on the flare active region, and it is derived from the
potential field source surface (PFSS) extrapolation
\cite{Schrijver03}. The flare active region is filled with a number
of magnetic field lines, including the open and closed lines, as
indicated by the purple and white lines, respectively. Also, the
flare area and the sunspot umbra are linked by some magnetic field
lines, providing an opportunity for the propagation of slow waves
from sunspot umbras to the flare area. This model was first
presented by \citeA{Nakariakov11}, who proposed that the QPPs of
double-ribbon flares could be explained by propagating slow waves.
Then, it is demonstrated by studying the relationship between HXR
pulses and footpoint sources in double-ribbon flares
\cite{Inglis12}. And the slow wave has been used to explain
the 3-min QPP in HXR, microwave, UV, and Ly$\alpha$ emissions
\cite<cf.>{Li24b}, similar to the 3-min period in our case.

\begin{figure}
\centering
\includegraphics[width=\textwidth]{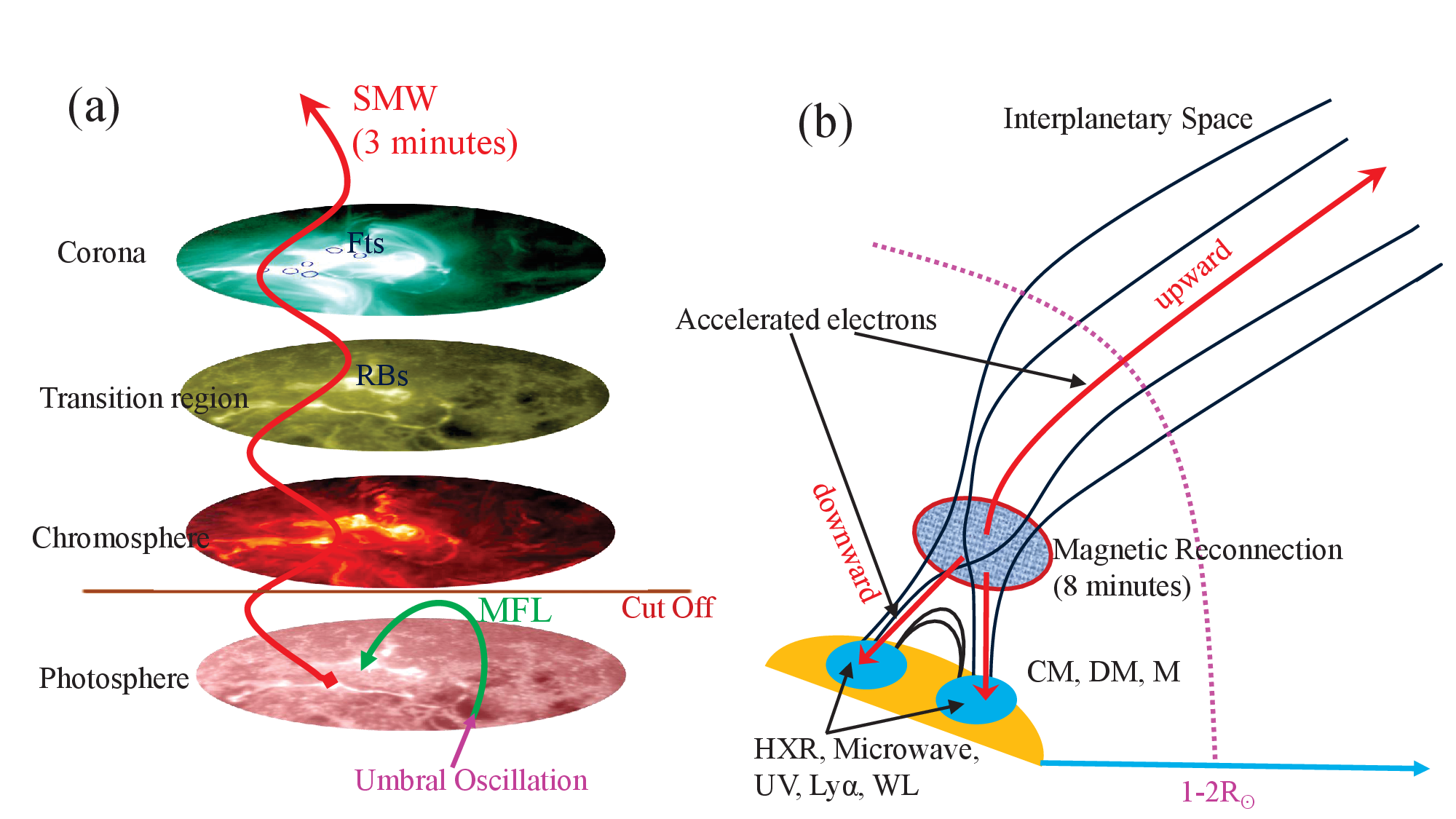}
\caption{Cartoon images for illustrating the origin of flare QPPs.
(a): The 3-min period is modulated by the slow
magnetoacoustic wave (SMW), which is originating from the umbral
oscillation in the sunspot. It propagates along the magnetic field
line (MFL) to the flare area, i.e., two ribbons (RBs) or double
footpoints (Fts). Here, the low-frequency cutoff prevents
long-period wave propagating from the photosphere to the upper
atmosphere. (b): The 8-min period is possibly triggered
by the repeated magnetic reconnection, which may periodically
accelerate non-thermal electrons. In this process, double footpoints
are formed in the HXR/microwave channels, two ribbons or some
kernels are generated in the UV, Ly$\alpha$, and white-light (WL)
emissions, and a group of radio bursts can be seen in various
frequencies of centimeter (CM), decimeter (DM) and meter(M) regimes,
or the low frequency in the interplanetary space. \label{cart}}
\end{figure}

Figures~\ref{pow}~(c1) and (d1) show the UV/EUV sub-maps in
wavelengths of AIA~1600 and 94~{\AA} during the X4.0 flare. Double
flare ribbons can be seen in the low-temperature waveband of
AIA~1600~{\AA}, and hot flare loops are observed in the
high-temperature waveband of AIA~94~{\AA}, as marked by the cyan and
tomato contours. Panels~(c2) and (c3) present the spatial
distributions of normalized Fourier power for the AIA~1600~{\AA}
data, and they are averaged among 2-4 and 7-9~minutes during
06:15-07:15~UT. We can find that the Fourier power spectra are
significantly enhanced in the double flare ribbons, including the
spectral components at 2-4 and 7-9 minutes. Panels~(d2) and (d3)
shows the spatial distributions of normalized Fourier power for the
AIA~94~{\AA} data, which are averaged among 2-4 and 13-15~minutes
during the same time interval. The spectral component at 2-4~minutes
appears to enhance at double flare ribbons or footpoints, while that
at 13-15~minutes appears at the hot flare loops. The spectral
component at 2-4~minutes is similar to the quasi-period at
3~minutes, and it can be simultaneously observed in wavebands of
HXR, radio, EUV/UV, and Ly$\alpha$ at double ribbons or footpoints.
Moreover, the similar period of 3~minutes is detected in the
adjacent sunspot umbras, and the flare ribbons and the sunspot
umbras are connected by some magnetic field lines. All those
observations support that the 3-min QPP is most likely modulated by
the slow magnetoacoustic wave leaking from the sunspot umbra
\cite<i.e.,>{Yuan11,Sych09,Inglis12,Kumar16,Li24b}. The idea is that
the slow magnetoacoustic wave at about 3~minutes is always existent
in the sunspot umbras, it originates from the photospheric umbra,
and then propagates along the magnetic field lines, causing the
flare QPP at about 3~minutes in the outer solar atmosphere
\cite<i.e.,>{Nakariakov11,Yuan14,Yuan15,Sych24}, as illustrated in
Figure~\ref{cart}~(a). The spectral component at 13-15~minutes is
similar to the quasi-period at 14~minutes seen in wavelengths of SXR
and high-temperature EUV, which may be associated with the periodic
variation of high-temperature plasmas that is heated by the
loop-loop interaction \cite{Li24a}, as the 14-min period is mainly
located in hot flare loops. We also note that the three main peaks
seen in the GOES flux originate from the same active region however
they appear to originate from different parts of the active region,
which are linked by some interaction between the magnetic fields of
loop systems in the active region. This also confirms that the
temperature evolution at a period of 14~minutes is attributed to the
loop-loop interaction. Moreover, the 14-min period can be
only detected in wavebands of SXR and high-temperature EUV, implying
that it depends on temperature.

\begin{figure}
\centering
\includegraphics[width=\textwidth]{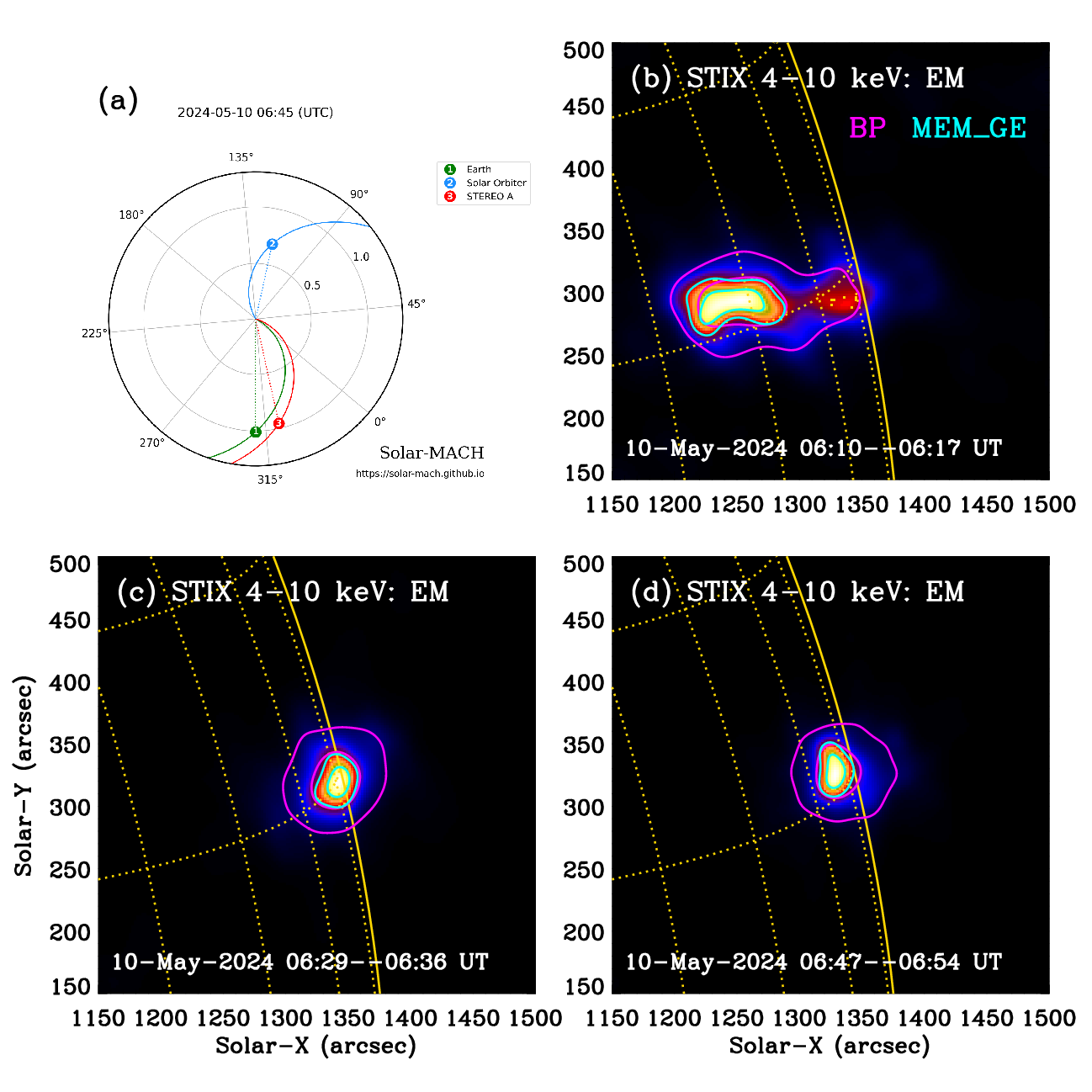}
\caption{(a): Sketch plot of the spatial locations of Solar Orbiter
and STEREO\_A and their connections with the Sun and Earth. (b)
Reconstructed STIX images at 4-10~keV using three algorithms (EM,
BP, and MEM\_GE). The contour levels are set at 50\% and 80\%. The
dotted gold lines represent latitude-longitude grids, and the solid
gold line marks the solar limb. \label{stix}}
\end{figure}

Figure~\ref{stix}~(a) shows the spatial location of the Solar
Orbiter with respect to Earth at 06:45~UT on 2024 May 10. STIX was
located at about 166.9 degree west in solar longitude from the
Sun-Earth line, their distance was about 0.69 AU. Thus, STIX
observed the Sun at a different point, comparing the Earth's
perspective. Panels~(b)-(d) presents the SXR images at three time
intervals, corresponding to the three pulsations in the STIX light
curve. The SXR images at 4-10~keV are reconstructed from the STIX
pixelated science data in the STIX Aspect System \cite{Warmuth20}.
For validating the source location of the reconstructed SXR
emissions, three image reconstruction methods that have been
implemented in the STIX data analysis software are used, that is,
the expectation-maximization algorithm (EM), the back-projection
method (BP), and the maximum entropy method (MEM\_GE). The source
locations of the second and third pulsations match with each other,
but they are far away from that of the first pulsations. Moreover,
the first pulsation locates inside the solar limb, so it was not
observed by GOES and SDO/AIA in the Earth measurement. Thus, the
quasi-period at about 18~minutes is only observed in the channel of
STIX~4-10~keV, which could be attributed to the intermittent
energy-release processes of different solar flares. That is, these
three pulsations are different in the space, but they are successive
and repetitive in the term of time series. Therefore, we still
termed it as the 18-min period, since they are flare emissions with
similar time intervals between three successive pulses
\cite{Zimovets21}, although they are in fact different flares.

We wanted to discuss the generation mechanism of the quasi-period at
about 8~minutes. These pulsations can be simultaneously seen in
wavebands of HXR, radio, EUV/UV, white light, and Ly$\alpha$
emissions. Moreover, it can be detected in a broad range of radio
frequencies, ranging from the centimeter through the decimeter to
the meter-decameter regimes. All these observations suggest that the
8-min period could be associated with nonthermal electrons that are
periodically accelerated by intermittent magnetic reconnections,
i.e., the periodic regimes of repetitive magnetic reconnection. The
idea is that the flare radiation in wavelengths of HXR, radio, white
light, and Ly$\alpha$ are usually related to accelerated electrons
via the well-known magnetic reconnection \cite{Priest02,Lin03}.
Therefore, the nonthermal electrons can be periodically accelerated
during the X4.0 flare, causing the 8-min QPP simultaneously in
wavelengths of HXR, radio, white light, and Ly$\alpha$ emissions, as
illustrated in Figure~\ref{cart}~(b). This is also consistent with
the white-light QPP reported in an X8.2 flare \cite{Zhao21}. On the
other hand, \citeA{Li24c} reported a white-light QPP at the period
of about 8.5~minutes, but it was only observed in the white-light
emission. This is obviously different from the 8-min period observed
in multiple wavelengths in our case. Therefore, the 8-min QPP can
not be modulated by the slow-mode magnetoacoustic gravity (MAG) wave
originating from the sunspot penumbra. If the slow-mode MAG wave at
the period of about 8~minutes is originated in the solar lower
atmosphere, then it cannot propagate upwardly into the solar upper
atmosphere, due to the low-frequency cutoff in the photosphere
\cite{Sych09,Su13,Yuan14,Kumar16}. That is, only the 3-min period
wave can propagate upwardly to the solar outer atmosphere, and the
quasi-period at about 3~minutes seen in double ribbons can well be
explained by the slow magnetoacoustic wave, as it can be
simultaneously observed in multi-height solar atmospheres, as shown
in Figure~\ref{cart}~(a). Moreover, the 3-min period can be
simultaneously seen in multiple AIA wavelengths, e.g., high- and
low-temperature channels, suggesting that the slow wave is
independent of temperature, which is similar to our previous
findings \cite{Li24b}.

\subsection{Coronal seismology and scaling laws}
Given the wide range of wavelengths considered, the flare
QPPs at multiple periods can be applied for the coronal seismology.
That is, the multi-period QPPs are useful for diagnosing the key
parameters of the flaring core, since they carry the time
characteristics of flare radiation \cite{Pugh19,Yuan19}. The flare
QPPs could be a novel tool to diagnose the stellar atmospheres if
their generation mechanisms are justified, because the spatial
resolution of a remote star is difficult to achieve
\cite{Zimovets21,Kowalski24}. In our case, the 2-min QPP is
explained by the periodic magnetic reconnection. The idea is that
nonthermal electrons are repeatedly accelerated by periodic magnetic
reconnections, which was predicted by the numerical simulation
\cite{Takasao16}. However, the periodic magnetic reconnection is not
detected directly due to the current instrumentation, that is,
coronal magnetic fields can not be measured directly. The QPP
feature observed in wavelengths of HXR and radio/microwave could be
associated with accelerated electrons generated by quasi-periodic
magnetic reconnections, such process has been discussed detailed in
previous observations \cite{Wu16,Li21,Kumar25}. The 8-min QPP can
also be interpreted by the periodic magnetic reconnection, since it
is simultaneously observed in wavelengths of HXR, radio/microwave,
Ly$\alpha$, and white light. This feature also suggests that the
nonthermal electrons may only precipitate into the lower
chromosphere or the upper photosphere, where is the main source
region of white-light emissions. Thus, the X4.0 flare belongs to
`Type I WLFs', as were categorized by \citeA{Fang95}. The 3-min
period is simultaneously observed at the flare site and the nearby
sunspot umbra, and they are connected by the closed magnetic field
lines. So, it is possibly regarded as the leakage of sunspot
oscillation in the form of slow MHD waves, which propagates from the
sunspot umbra to the flare site along the closed magnetic loop, as
shown in Figure~\ref{cart}~(a).

The coronal seismology aims to estimate the local plasma and
magnetic properties using observed wave characteristics and their
underlying mechanisms. According to the MHD waves at kink
\cite{Nakariakov21,Gao24} and slow modes \cite{Wang21,Kolotkov22},
some plasma and magnetic properties in the corona can be estimated
from Equations~\ref{eqs1} and \ref{eqs2}, respectively. In our case,
both the 2-min and 8-min QPPs can be explained by the mechanism
based upon a repetitive regime of magnetic reconnection, which might
be modulated by the resonances of MHD waves. The 8-min QPP appears
during the X4.0 flare, which shows a hot plasma loop
(L1~$\approx$~92~Mm), as indicated by the red arrow in
Figure~\ref{img1}~(b2). The phase speed ($v_{\rm p}$) is estimated
to about 400~km~s$^{-1}$, which is less than the local sound speed
($v_{\rm s}\approx$~480~km~s$^{-1}$) in the solar corona at a high
temperate ($T$) of about 10~MK. The polytropic (or effective
adiabatic) index ($\alpha$) can be estimated to about 1.16,
consistent with that measured by \citeA{Van11}. By using the
relation between $v_{\rm p}/v_{\rm s}$ and thermal ratio ($d$) given
by \citeA{Wang21}, the effective thermal ratio $d_{\rm e}$ could be
estimated to about 0.18, the actual thermal ratio ($d_{\rm 0}$) is
about 0.016, and the result of $d_{\rm e}$/$d_{\rm 0}$ is about
11.25 in the flare loop. All these facts suggest that the 8-min
period is possibly modulated by the slow-mode wave. On the other
hand, a hot plasma loop (L2~$\approx$~63~Mm) is seen during the
pre-flare phase (Figure~\ref{pre}~c). The phase speed is about
1050~km~s$^{-1}$, which is in the range of kink speeds
\cite{Nakariakov20,Nakariakov21}. Therefore, the 2-min period is
likely to be modulated by the kink wave, and the local Alfv\'{e}n
speed ($v_{\rm A}$) and magnetic field strength ($B$) are estimated
to about 750~km~s$^{-1}$ and 40~G in the hot plasma loop, agreeing
with our previous measurements in flare loops \cite{Li17,Li18}. It
should be pointed out that all those estimations are in the corona,
i.e., hot plasma loops. The plasma and magnetic properties in lower
atmospheres are difficult to be measured, largely due to their
complex magnetic fields and high densities.

\begin{equation}
  v_{\rm p} = \frac{2L}{P}, \quad
  v_{\rm A} = v_{\rm p}~\sqrt{\frac{1+r_{\rho}}{2}}, \quad
  B  \approx  v_{\rm A}~\sqrt{\mu_0~\rho}.
  \label{eqs1}
\end{equation}
\noindent Where, $r_{\rho}$ is the density ratio between external
and internal flare loop, $\mu_0$ is the magnetic permittivity in
vacuum, and $\rho$ is the density of hot loop.

\begin{equation}
  v_{\rm s} \approx 152~\sqrt{T({\rm MK})}, \quad
  v_{\rm p} \approx \sqrt{\frac{\alpha}{\gamma}}~v_{\rm s}, \quad
  d_{\rm 0} = 4.93~(\frac{T^{3/2}}{n~P}).
  \label{eqs2}
\end{equation}
\noindent Here, $\gamma$ is the adiabatic index, and $n$ is the
number density of flare loop. $T({\rm MK})$ refers to that the loop
temperature in the unit of MK.

The flare QPPs at periods of 3 and 8 minutes are both
associated with slow-mode waves, but their generation mechanisms are
clearly different. In the coronal environment, large-amplitude QPPs
(i.e., $>$20\%) could be triggered by slow magnetoacoustic waves
when the duration of the heat pulse ($\triangle t_{\rm H}$) is
shorter than the loop sound crossing time ($\tau_{\rm s}$)
\cite{Reale19,Wang21}, as shown in Equation~\ref{eq2}. In our case,
$\tau_{\rm s}$ can be estimated to about 460~s at the X4.0 flare
maximum. Under the assumption of $\triangle t_{\rm H} \simeq P$, the
3-min QPP may be directly modulated by the slow magnetoacoustic wave
leaking from sunspot umbrae, as is illuminated in
Figure~\ref{cart}~(a). Conversely, the 8-min QPP is generated by the
repetitive magnetic reconnection that is triggered by the slow-mode
wave in the solar corona, that is, the periodicity of magnetic
reconnection is modulated by the slow-mode wave at a period of about
8~minutes, as shown in Figure~\ref{cart}~(b). Similarly, the 2-min
QPP is generated by the periodic magnetic reconnections that is
triggered by the kink wave.

\begin{equation}
  \triangle t_{\rm H} < \tau_{\rm s} \sim 5~\frac{L({\rm Mm})}{\sqrt{0.1~T({\rm MK})}}.
  \label{eq2}
\end{equation}

In order to compare with the statistically established
scaling laws for flare QPPs \cite{Hayes20}, we perform the scaling
law between the QPP period (P) and the flare duration ($\tau$) with
the relationship described in Equation~\ref{eq1}. $\tau$ is
determined by the time interval of the start and stop times of the
X4.0 flare that was defined in the GOES catalog, and it is 2760~s. C
is a constant, which is about 0.5 given by \citeA{Hayes20}. Then,
the QPP period is estimated to be (1.68$\pm$0.01)~minutes, which is
clearly different from the multiple periods detected in this study.
This is mainly because that the scaling laws presented by
\citeA{Hayes20} only searched for periodicity in the timescale of
6$-$300~s in the GOES SXR channel, but the multiple periods in this
study are found in a wide range of wavelengths. Therefore, the
scaling laws for flare QPPs should be updated if we considering a
wide wavelength range and a longer period range.

\begin{equation}
\centering
 P~\approx~C\tau^{0.67 \pm 0.03}.
\label{eq1}
\end{equation}

\section{Summary}
Based on the combined solar telescopes, i.e., ASO-S/HXI, ASO-S/LST,
SDO/AIA, SDO/HMI, SDO/EVE, NVST, CHASE, SUTRI, GECAM, STIX, LYRA,
GOES, CBS, DART, SWAVES, and NoRP, we explored flare QPPs at five
quasi-periods in multiple wavebands, we also tried to explore their
generation mechanisms. The main conclusions are summarized as
follows:

\begin{enumerate}
\item The multi-periodic pulsations are detected during an X4.0 flare in
various wavebands, i.e., at least five periods. The X4.0 flare could
be regarded as a WLF, as it is simultaneously brightening in
wavebands of WST~3600~{\AA}, HMI continuum near 6173~{\AA}, and NVST
TiO 7058~{\AA}. It belongs to the `Type I WLFs'. This is the first
report of the flare QPP that is seen in the channel of
TiO~7058~{\AA}.

\item A quasi-period at about 2~minutes is observed in HXR and
radio (i.e., centimeter and decimeter) emissions during the flare
precursor, which might be associated with the periodic magnetic
reconnection, and it could be regarded as a precursory indicator of
the powerful flare.

\item A quasi-period at about 3~minutes is simultaneously detected in
channels of HXR, radio, and EUV/UV at double ribbons or footpoints.
It is independent of temperature and could be directly modulated by the slow
magnetoacoustic wave leaking from sunspot umbrae.

\item A quasi-period at about 8~minutes is simultaneously detected in
wavebands of HXR, radio, EUV/UV, white light and Ly$\alpha$ during
the X4.0 flare. It is probably triggered by quasi-periodic regimes
of magnetic reconnection, which can periodically accelerate
nonthermal electrons.

\item A quasi-period at about 14~minutes is found in wavelengths
of SXR and high-temperature EUV from the precursor to the X4.0
flare. It depends on temperature and might be associated with the
temperature fluctuation caused by loop-loop interaction.

\item A quasi-period at about 18~minutes is seen in
STIX~4-10~keV. It could be regarded as three successive and repeated
energy-release processes in different flare regions.

\end{enumerate}

\section*{Open Research Section}
Publicly available data sets were analyzed in this study. They can
be found here: ~~http://aso-s.pmo.ac.cn/sodc/dataArchive.jsp,
~~https://sun10.bao.ac.cn/SUTRI/,   \\
~~https://ssdc.nju.edu.cn/NdchaseSatellite,
~~http://jsoc.stanford.edu/, \\
~~http://www.solarmonitor.org/.

\acknowledgments This work is funded by the National Key R\&D
Program of China 2021YFA1600502 (2021YFA1600500), and 2022YFF0503002
(2022YFF0503000). This work is also supported by the Strategic
Priority Research Program of the Chinese Academy of Sciences, Grant
No. XDB0560000, NSFC under grants 12473059. D.~Li is also supported
by the Specialized Research Fund for State Key Laboratories. We acknowledge the use of data from the Chinese MeridianProject.
ASO-S mission is supported by the Strategic Priority Research Program on
Space Science, the Chinese Academy of Sciences, Grant No.
XDA15320000. SUTRI is a collaborative project conducted by the
National Astronomical Observatories of CAS, Peking University,
Tongji University, Xi'an Institute of Optics and Precision Mechanics
of CAS and the Innovation Academy for Microsatellites of CAS. The
CHASE mission is supported by China National Space Administration
(CNSA). CBS consists of four subsystems is operated by SDU. DART is
an interferometric imaging telescope supported by Chinese initiative
Meridian Project phase 2. SDO is NASA's first mission in the Living
with a Star program. The STIX instrument is an international
collaboration between Switzerland, Poland, France, Czech Republic,
Germany, Austria, Ireland, and Italy.

\bibliography{my_references}

\end{document}